\documentclass[12pt, hidelinks]{article}
\usepackage[twoside=true, margin=1in]{geometry}

\usepackage[english]{babel}
\usepackage{bm}
\usepackage{import}
\usepackage[utf8]{inputenc}
\usepackage[english]{babel}
\usepackage{amsmath}
\usepackage{verbatim}
\usepackage{bbm}
\usepackage{dsfont}
\usepackage{amsthm}
\usepackage{amssymb}
\usepackage{enumitem}
\usepackage{mathrsfs}
\usepackage{tikz}
\usepackage{tikzsymbols}
\usepackage{soul}
\usepackage[numbers]{natbib}
\usepackage{hyperref}
\usepackage{xcolor}
\usepackage{color}
\usepackage{caption,graphicx,newfloat}
\graphicspath{ {images/} }
\usepackage{booktabs} % For formal tables
\usepackage[ruled]{algorithm2e} % For algorithms
\usepackage{tcolorbox}
\theoremstyle{plain}

\newtheorem{assumption}{Assumption}[section]

\newtheorem{claim}{Claim}[section]

\theoremstyle{remark}

\title{A Curationary Tale: \\ 
Logarithmic Regret in DeFi Lending via Dynamic Pricing}
\author{Tarun Chitra\\ 
        Gauntlet \\
        \texttt{\small tarun@gauntlet.xyz} 
       }
\date{\today}
\newcommand{\ones}{\mathbf 1}
\newcommand{\reals}{{\mbox{\bf R}}}

\newcommand{\naturals}{{\mbox{\bf N}}}

\newcommand{\Expect}{\mathop{\bf E{}}}

\newcommand{\Prob}{\mathop{\bf Prob}}
\newcommand{\argmin}{\mathop{\rm argmin}}
\newcommand{\argmax}{\mathop{\rm argmax}}
\newcommand{\cf}{{\it cf.}}
\newcommand{\eg}{{\it e.g.}}
\newcommand{\ie}{{\it i.e.}}

\newcommand{\BEAS}{\begin{eqnarray*}}
\newcommand{\EEAS}{\end{eqnarray*}}
\newcommand{\BEA}{\begin{eqnarray}}
\newcommand{\EEA}{\end{eqnarray}}
\newcommand{\BEQ}{\begin{equation}}
\newcommand{\EEQ}{\end{equation}}
\newcommand{\BIT}{\begin{itemize}}
\newcommand{\EIT}{\end{itemize}}

\begin{document}
\maketitle

\begin{abstract}
    Lending within decentralized finance (DeFi) has facilitated over~\$100 billion of loans since 2020.
    A long-standing inefficiency in DeFi lending protocols such as Aave is the use of static pricing mechanisms for loans.
    These mechanisms have been shown to maximize neither welfare nor revenue for participants in DeFi lending protocols.
    Recently, adaptive supply models pioneered by Morpho and Euler have become a popular means of dynamic pricing for loans.
    This pricing is facilitated by agents known as curators, who bid to match supply and demand.
    We construct and analyze an online learning model for static and dynamic pricing models within DeFi lending.
    We show that when loans are small and have a short duration relative to an observation time $T$, adaptive supply models achieve $O(\log T)$ regret, while static models cannot achieve better than $\Omega(\sqrt{T})$ regret.
    We then study competitive behavior between curators, demonstrating that adaptive supply mechanisms maximize revenue and welfare for both borrowers and lenders.  
\end{abstract}

\section{Introduction}
Overcollateralized lending is one of the most popular decentralized applications hosted on blockchains.
Since the 2019 introduction of the first overcollateralized lending protocol --- Compound Finance --- there have been over~\$100 billion in loans facilitated on-chain.
These protocols match suppliers who aim to earn a return on their assets with borrowers who prefer to borrow rather than sell a particular cryptocurrency.
One key feature of these protocols is that they operate in permissionless and adversarial environments.
This means that credit generically needs to be fully secured as opposed to lending based on, \eg, credit scores.
Lending protocols ensure solvency via a combination of liquidation schemes to sell collateral of defaulted loans and dynamic adjustment of parameters, such as the loan-to-value ratio.

Overcollateralized loans are often viewed as a capital inefficient form of lending as they require the borrower to utilize collateral worth more than what they borrow.
As such, there has been a natural evolution of lending protocols towards those that optimize efficiency in terms parameters such as collateral requirements and interest rates charged.
This optimization has led to numerous designs for lending protocols that specialize to different types of borrowers.
By offering loans tailored towards a particular borrower demographic, protocols can adjust capital requirements and interest rates in a real-time fashion in response to large market changes.
This explains the dramatic increase in the number of application-specific lending protocols such as Jupiter's JLP~\cite{chitra2025perpetual}, Pendle interest rate swaps~\cite{angeris2023replicating, pendle}, and Hyperdrive's variance swap-like protocol~\cite{rhea2024hyperdrive}.

\paragraph{First and Second Generation Lending Protocols.}
There are numerous axes under which lending protocols have been improved since 2019.
These dimensions include market isolation, dynamic collateral requirements, dynamic pricing, and supply reallocation.
We will first briefly cover how different protocols improved each of these areas.

The first overcollateralized lending protocols, such as Compound V2 and Aave V2, allowed for many-to-many borrowing.
This meant that depositors could deposit multiple types of collateral, such as USDC, ETH and WBTC, and borrow multiple types of borrowed assets against this portfolio.
In practice, this forces loan-to-value ratios to be much more conservative since the LTV has to take into account the overall correlation between the collateral and borrower portfolios~\cite{fuAave, kao2020analysis}.
In particular, the liquidity of worst quality collateral asset would gate how aggressive one could be with loan-to-value ratios.

As an improvement on this initial model, protocols such as Aave V3 and Compound V3 allowed for isolation modes.
We term lending protocols with isolation modes as second-generation lending protocols.
These isolation modes allowed some subset of the supplied assets to be lent only against a particular collateral.
This, in turn, allowed for higher loan-to-value ratios to be offered, due to the decreased liquidation risk of particular pairs of collateral and borrow assets.
Isolation modes also became extremely popular for borrowing yield-bearing assets against yield-free assets, allowing users to execute delta hedging strategies with lending protocols.
These isolated markets grew to more than \$5 billion between 2021 and 2025, showing the dramatic growth that a relatively simple mechanistic change can produce.

We note that both the first generation and isolated pools utilized static pricing to construct interest rate curves.
This means that a fixed interest rate function $I(\kappa, D, S)$ is chosen ahead of time, where $\kappa$ is a set of hyperparameters, $D$ is the aggregate loan demand, and $S$ is the asset supplied to be lent.
Most protocols utilize governance mechanisms, such as Compound Governance~\cite{papangelou2023exploring, saengchote2023decentralized}, to infrequently adjust the parameters $\kappa$.
However, in between governance votes, the parameters are static, leading to fixed pricing.

\paragraph{Third-Generation Lending Protocols.}
While isolation allows for increased capital efficiency, it creates another problem, liquidity fragmentation.
Capital that is allocated to an isolated market cannot be used within another market, so when an isolated market has low utilization, lending protocols lose out on revenue that they could have earned by allocating the same capital to higher utilization markets.
The first two generations of lending protocols updated reallocation parameters, such as supply caps and borrow caps, through governance, which occurs at a low frequency.

Moreover, it has been empirically demonstrated that lenders in first-generation and isolated lending protocols have low elasticity to interest rates, as they treat the pool as a passive investment~\cite{chaudhary2023interest, aave-ir-curve-changes, }.
This means that while lenders could reallocate their pooled assets on their own to correct for the fragmentation issue, in practice, they are unable to do this due to a lack of technical and/or financial sophistication.

This inefficiency led to the development of competitive, decentralized mechanisms to reallocate supply across markets and adjust prices accordingly.
The third generation over-collateralized lending markets, pioneered by Morpho~\cite{metamorpho}, have third parties competing to perform supply reallocations to optimize utilization of supplied assets.
These markets have third-parties known as \emph{curators} who rebalance supply in lending markets.
Curators compete for capital from lenders by offering a combination of matching supplied capital with borrowers at competitive prices (by reallocating supply across markets) and risk management.

The curator fulfills the dual mandate of choosing dynamic prices (\ie~interest rates) as loans arrive, while also allocating in a manner to reduce defaults.
These prices are quoted based on how much supply a curator adds to a given loan (or pool of loans).
This notion of dynamic pricing, where the price paid changes depending on how curators allocate supply across markets, is distinct from prior lending models.\footnote{Existing lending protocols, such as Aave, are adding features such as the so-called liquidity premia~\cite{aavev4-risk-premia}, that attempting to emulate the supply rebalancing inherent in the curator model.}

There are numerous curation-based lending protocols, and another prominent example is Euler Finance.
The curation markets within Morpho have grown from less than 0. 5\% of the lending market to nearly 12\% of assets in decentralized lending between 2024 and 2025.
Currently, in March 2025, Morpho has over~\$5 billion~\cite{defillama-morpho} in assets while those of Euler have grown to over~\$700 million~\cite{defillama-euler}. 

We note that different vaults generally have varied credit risk.
For instance, Euler and Morpho differ in the way they underwrite credit risk, as Euler allows for multi-collateral borrows~\cite{euler-earn-wp} and utilizes governance to classify assets into three tiers: isolated, cross, or collateral.
Isolated assets are those that only be used as a borrowable asset in isolation, cross represents assets that can be borrowed along with other assets and, collateral assets can be used anywhere in the protocol.

Morpho, on the other hand, has only completely isolated markets that allow for single collateral, single borrowable asset loans.
These markets can be aggregated into ``meta" vaults, which are managed by curators.
In particular, curators take a borrowable asset (that is supplied with the aim of earning yield) and allocate it across multiple isolated markets.
Our single borrowable asset~\S\ref{sec:single-dimensional-model} represents Morpho's isolated markets, whereas our multiple borrowable asset model~\S\ref{sec:multidimensional-model} represents both Metamorpho and Euler vaults.\footnote{We note, however, that we assume that each vault represents a unique set of collateral and borrowable assets; we leave a description of vault non-fungibility for future work (c.f.~\cite{diamandis2025market})}

\paragraph{Lending Protocols as Online Learning Algorithms.}
One key question all DeFi lending protocols is how to optimally choose interest rates.
First generation DeFi protocols generally parametrized interest rate curves as piecewise linear functions of the utilization, $U = \frac{D}{S}$, which is the ratio of demand to supply.
The slope(s) of such curves represents elasticity of interest rates to changes in demand.
Prior work has focused on modeling the interest rate optimization problem from the perspective of either stochastic control theory~\cite{bastankhah2024thinking, bastankhah2024agilerate, bertucci2024agents, baude2025optimal} or equilibrium, mean-field models~\cite{cohen2023economics, rivera2023equilibrium}.
The commonality that these models have is that they model supply and demand as stochastic processes and compute continuous-time limits for these processes.
These limits either give equilibrium conditions in the limit of a large number of participants or provide a limiting stochastic differential equation that can be analyzed numerically.

However, DeFi lending is in many ways inherently discrete.
Blockchains ensure that these systems operate on discrete clocks so that loan durations are integral.
Moreover, the stochastic processes for supply and demand have constraints placed upon them that ensure that they are always bounded.
We first argue that online learning and online convex optimization is a more natural framework for analyzing all generations of lending protocols.
Given the success of online learning in achieving super human performance in numerous tasks such as Poker~\cite{brown2019superhuman}, Diplomacy~\cite{meta2022human}, and market making~\cite{cesa2024market, abernethy2013adaptive}, it is natural to analyze it within competitive lending markets.

Note that the lending demand in a DeFi protocol arrives sequentially.
One can view this as a sequence of loans $\ell_1, \ldots, \ell_k$ that arrive sequentially.
At each point in time, a lending protocol quotes a price $p_i$ to loan $i$.
Static pricing models, such as those used in Aave or Compound, make $p_i$ a deterministic function of $\ell_1, \ldots, \ell_i$.
Note that the price $p_i$ does not depend on loans in the future $\ell_j, j >i$.

On the other hand, curatorial models have $p_i$ priced by curators to optimize their revenue, who can be viewed as actors choosing prices strategically and in response to competition from other curators.
This implies that the price $p_i$ will be dynamic, depending on the composition of liquidity available to each curator and their bidding strategy.\footnote{Although we don't study this explicitly, the curation model can be seen as a particular type of all-pay auction.
One can imagine extending the intents model of~\cite{chitra2024analysis} and the all-pay model for ZK auctions of~\cite{succinct-wp} to describe curation.}
We note that this dynamic pricing, much like the static pricing, can only depend on loans that have arrived so far (\ie~$p_i$ cannot depend on $\ell_j$ for $j > i$).

As is common in online learning, we can consider an optimal benchmark $p^{\star}_i$, which are the batch prices.
These prices are chosen to optimize an objective function of the prices and allow the prices $p^{\star}_i$ to depend on the entire batch of loans.
Given the optimal prices, we can measure \emph{(external) regret}, which is the difference between the objective function at the optimal value and the expected value of the objective function of a given online algorithm.
In this paper, we will be concerned with protocol revenue as the objective function.
The asymptotics of the regret bound in the number of loans and their duration provide a means for comparing different online algorithms, such as those used in Aave and Morpho.
If the regret of an algorithm is sublinear, then classical results demonstrate that we converge to a correlated equilibrium between borrowers and lenders~\cite{foster1997calibrated, foster1999regret}.

% We consider two types of objective functions: revenue and welfare.
% Revenue represents the total income received by the lenders in a DeFi lending protocol and the prices $p^{\star}_i$ correspond to the best prices one could choose in hindsight, had they known the demanded loans ahead of time.
% Welfare incorporates a notion of intrinsic value for a loan from borrower --- for instance, two different borrowers might value a loan of a particular size differently.
% Our goal is to compare how the revenue or welfare generated by the online algorithms for interest rate computations compare to that generated by the optimal prices. 

\paragraph{Lending and Intents.}
Another DeFi market that resembles third generation lending protocols is the so-called intents market for decentralized trading~\cite{chitra2024analysis, yuki-paper}.
In this market, actors known as solvers bid in a dutch auction to provide users who fulfill orders for users who want to swap assets.
Much like competing curators, these actors compete with one another to provide better prices to users versus passive liquidity.
The prices bid by solvers are compared to a static price quoted by a constant function market maker (CFMM) and if no price improves upon the CFMM's price, the trade is executed against the CFMM.

One can view intents as loans with zero duration, which is a single-shot learning problem (versus a multiperiod learning problem).
Intents are generally simpler than loans and can be analyzed explicitly given parametric information about the size of the trades involved.
However, from the analogy of loans as single-round intents, one can find analogues of known results about intents that hold for loans.

\paragraph{Related Work.}
Dynamic pricing and inventory management have been studied in a number of different fields.
One work that is similar to our methodology is Besbes, et. al~\cite{besbes2009dynamic}, where there is an unknown demand function and a seller is trying to dynamically optimize their price.
They find polylogarithmic regret bounds similar to those presented here, albeit with different exponents.
Another work that analyzes regret in dynamic pricing is Babaioff, et. al~\cite{babaioff2015dynamic}, which considers a generalization of Besbes's model with restricted supply.
Both of these paper don't deal with inventory management, however, which is crucial within decentralized finance.
As our model allows for elastic supply, we have significantly tighter regret bounds.
We also find similarities to our methodology within the inventory management literature, especially with regards to solutions to the newsvendor problem~\cite{qin2011newsvendor, gallego1994optimal}.
Finally, we note that the curator market resembles the online bipartite matching market with reusable resources, which is common within ride-sharing~\cite{dickerson2021allocation}.

\paragraph{This Paper.}
We first formalize an online learning model for the interest rates of the lending protocol.
This requires us to define the revenue generated by a lending protocol in a generic manner that handles first, second, and third generation lending protocols.
We first consider loans of a single borrowable asset against a single collateral asset.
The main components of this model involve a stochastic process for loans, a price quoting algorithm, and a notion of loan duration.
Note that the price quoting algorithm can be centralized and static, like Aave, or decentralized and dynamic, like Morpho.
We then generalize this model to multiple collateral and borrowable assets. 

Using this model, we first analyze fixed rate lending, which corresponds to a price $p_i$ for each loan and charging that loan that price throughout the loan's duration.
In the fixed rate model, we are able to show that the static pricing models used by Aave and Compound have regret $\Omega(T)$.
We do this by constructing explicit worst case examples where their pricing algorithms have constant regret on every time step.
For the same example, we find that the Morpho model has sublinear regret.

We then show that if the stochastic process for the arriving loans chooses loans smaller than the assets held by curators and of shorter duration with high probability, then the regret of the Morpho model is $O(\log T)$.
The key insight in the proof is that strategic curators optimizing their own revenue will ensure that the curvature of the overall revenue function is large in order to maximize their own revenue.
Combining this with classical results on online learning~\cite{hazan2007logarithmic, zinkevich2003online} that regret scales as $O\left(\frac{1}{\text{curvature}} \cdot\log T\right)$ yields the result.
This demonstrates there is effectively a phase transition between Morpho and Aave under particular loan distributions.

We then analyze a model of variable rate lending where the interest rate charged for loan $j > i$ impacts the price loan $i$ pays.
This is equivalent to the model used by Aave and Morpho in practice, where the interest rate is variable and updated for all active borrowers.
While this adds the complexity of path dependence within the analysis, we demonstrate that one can bound the variable rate model's regret for Morpho as $O(\Expect[\tau] \log T)$, where $\Expect[\tau]$ is the expected loan duration.
As we found with the fixed rate model, Aave still has $\Omega(T)$ regret for their lending model, even with variable rates.
This implies that if the loan duration is sublinear in the variable rate model, then one can achieve sublinear regret.

\paragraph{Notation.}
\begin{itemize}
    \item $\Delta^n = \{ (w_1, \ldots, w_n) \in \reals^n_+ : \sum_i w_i = 1\}$
    \item For a natural number $k \in \naturals$, $[k] = \{1, 2, 3, \ldots, k\}$.
    \item We utilize the following Landau notations for functions $f : \reals \rightarrow \reals$:
    \begin{itemize}
        \item $f(n) \in O(g(n))$ if there exists $K > 0$ such that $f(n) \leq K g(n)$
        \item $f(n) \in o(g(n))$ if $\frac{f(n)}{g(n)} \rightarrow 0$ as $n \rightarrow \infty$
        \item $f(n) \in \Omega(g(n))$ if there exists $K > 0$ such that $f(n) \geq K g(n)$
    \end{itemize}
\end{itemize}

\section{Background on Online Learning}\label{sec:background}
In this section, we will provide a brief description of the mathematical framework for online learning and online convex optimization.
Our coverage will be brief and only touch on results that we utilize in the sequel.
We refer the reader to surveys~\cite{mcmahan2017survey, hoi2021online} and their references within for a more detailed description.

\paragraph{Online Learning and Optimization.}
Our setting starts with a set $K \subset \reals^n$ and an objective function $f : K \rightarrow \reals$.
Our goal is to find an optimum of $f$, $x^{\star} \in \argmin_{x \in K} f(x)$.
We will consider maxima given that our main objective in this paper is revenue maximization, but note that one can simply negate the objective function solve for minimization problems.
Generally, one can find an optimum using either global or local methods.
Global methods, such as a interior point methods~\cite[11.7]{boyd2004convex}, try to describe a simplified approximation of the set $\mathcal{A}$, where an exact solution can be found.
Local methods, on the other hand, involve sampling a set of points $x_1, \ldots, x_T \in \mathcal{A}$ and estimating the optimum given these points.
While local methods have worse theoretical guarantees, they have enjoyed far more success in practical usage due to their simplicity and computational tractability.

One specialized form of local optimization is online learning.
In online learning, an optimizer is given an arbitrary, unknown sequence of cost functions $f_1, \ldots, f_T$ and is tasked with choosing a sequence of points $x_1, \ldots, x_T$ in order to minimize $\sum_{t=1}^T f_t(x_t)$.
When the optimizer has to choose the point $x_t$, they are only allowed to use information about $f_s$ (and its subgradients) for $s \leq t$ and the previous chosen points $x_1, \ldots, x_{t-1}$.
We denote an online learning algorithm as $\mathcal{A}$, where the $t$th decision satisfies $x_t = \mathcal{A}(x_1, \ldots, x_{t-1})$.

As an example, suppose that $x_t \in \Delta^n$ represents a portfolio of assets.
At each time $t \in [T]$, an optimizer chooses $x_t$ and realizes loss $-\log(p_t^T x_t)$, where $p_t$ is the set of prices of the assets.
As the optimizer does not know the prices ahead of time, an online learning algorithm tries to maximize returns in the face of the uncertainty of price changes.

\paragraph{Regret and Dynamic Regret.}
To measure the performance of an online algorithm, a natural benchmark is to compare to the best decision that one would have chosen had they known the points $x_1, \ldots, x_T$ in advance.
One way of measuring this is the \emph{(External) Cumulative Regret}, which is defined as
\[
\mathsf{Regret}(\mathcal{A}, T) = \max_{x \in K} \sum_{t=1}^T f_t(x) - \sum_{t=1}^T f_t(x_t) 
\]
This can be viewed as measuring the difference between the best action that one could take in hindsight versus the actions chosen by the online algorithm $\mathcal{A}$.

We note that if the algorithm $\mathcal{A}$ is stochastic, one often refers to the expected regret, $\Expect[\mathsf{Regret}(\mathcal{A}, T)]$.
Often times, an algorithm $\mathcal{A}$ will choose a sampling procedure (\ie~stochastic gradient descent) to choose the point $x_t$ and we can view the points $x_1, \ldots, x_T$ as being sampled from a probability distribution $\pi_t$.
If this distribution is stationary and converges to an equilibrium distribution $\pi^{\star}$, \ie~$d_{TV}(\pi_t,\pi^{\star})\rightarrow 0$ as $t\rightarrow \infty$, then asymptotically, the expected regret is also stationary~\cite{zhao2020dynamic, zhang2017improved, zinkevich2003online}.

However, for non-stationary distributions, this definition of expected regret is insufficient.
For non-stationary environments, one defines the worst-case~\emph{dynamic regret} as~\cite{zhang2017improved}
\[
\mathsf{DRegret}(\mathcal{A}, T) =  \sum_{t=1}^T \max_{x\in K} f_t(x) - \sum_{t=1}^T f_t(x_t)
\]
This captures the optimum value for each update.
If the process is stationary, then the two notions of regret are equivalent; however, only dynamic regret is monotone and well-defined when the process isn't stationary~\cite{zhao2020dynamic}.

One goal of online learning research is so show that the average regret goes to zero as $T\rightarrow \infty$.
That is, one wants
\[
\frac{1}{T}\Expect[\mathsf{Regret}(\mathcal{A}, T)] \rightarrow 0
\]
Note that one can view the optimization problem for minimizing $\sum_t f_t$ as a game between a player who chooses $x_t$ and an adversary who chooses $f_t$.
Classical results~\cite{foster1997calibrated, foster1999regret} show that one converges to a correlated equilibrium as the average regret reaches zero.
This is the basis of many successful uses of online learning in practice, such as superhuman capabilities at Poker~\cite{brown2019superhuman} and Diplomacy~\cite{meta2022human}.

\paragraph{Competitive Ratios.}
Another measure of the performance of the online algorithm is the so-called~\emph{competitive ratio}~\cite{borodin2005online}.
An online algorithm $\mathcal{A}$ has a feasible competitive ratio of $\alpha$ if
\[
\sum_{t=1}^T f_t(x_t) \geq \alpha \sum_{t=1}^T \max_{x_t \in K}f_t(x_t)
\]
We define the competitive ratio $\mathsf{CR}(\mathcal{A}, T)$ as the supremum of all such $\alpha$, \ie~
\[
\mathsf{CR}(\mathcal{A}, T) = \sup\left\{ \alpha : \sum_{t=1}^T f_t(x_t) \geq \alpha \sum_{t=1}^T \max_{x_t \in K}f_t(x_t), x_i = \mathcal{A}(x_1, \ldots, x_{i-1})\right\}
\]
The competitive ratio is a multiplicative analogue of regret representing what percentage of the optimum value an online algorithm can get.
For example, if $\mathsf{CR}(\mathcal{A}, T) = O\left(\frac{1}{T}\right)$, then the online learning algorithm has a performance increasingly worse than optimal as $T \rightarrow \infty$.
On the other hand, if $\mathsf{CR}(\mathcal{A}, T) = \Omega(1)$, then the performance of the online algorithm does not degrade with time.

We note that there can often be a divergence between regret bounds and competitive ratios, so having explicit bounds on both quantities is useful for comparing different online algorithms~\cite{daniely2019competitive}.
For example, it is possible for an online algorithm to have $\mathsf{CR}(\mathcal{A}, T) \geq c > 0$, while also still having linear regret (\ie~$\frac{1}{T} \Expect[\mathsf{Regret}(\mathcal{A}, T)]\rightarrow c > 0$).
However, despite the poor regret bound, if $c$ is sufficiently large, it might be fine for a practical online algorithm to realize $c$% of the optimal value.
This phenomenon is often found in~\eg, online bipartite matching~\cite{devanur2013randomized, karp1990optimal}.

\paragraph{Known Bounds.}
If the functions $f_t$ and the set $K$ are convex, one can often bound the regret generically.
Recall that a set $K \subset \reals^n$ is convex if for all $t \in [0, 1]$ and points $x, y \in K$, $tx +(1-t)y \in K$.
Similarly, a function $f : K \rightarrow \reals^n$ is convex if for all $x, y \in K, t \in [0,1]$, we have $f(tx+(1-t)y) \leq t f(x) + (1-t)y$.
A function $f : K \rightarrow \reals^n$ has a subgradient $g$ at $x$ if $f(x) \geq f(y) + g^T(x-y)$ and we denote the set of all subgradients at $x \in K$ as $\partial f(x)$.
We say that a function $f : K \rightarrow \reals^n$ is $\mu$-strongly convex if for all $x \in K$ and all $g \in \partial f(x)$ we have
\[
f(y) \geq f(x) + g^T(y-x) + \frac{\mu}{2} \Vert y-x\Vert_2^2
\]
Finally, recall that a function $f : K \rightarrow \reals^n$ is a $L$-Lipschitz if for all $x, y \in K, |f(x)-f(y)| \leq L\Vert x - y\Vert_2$.

\noindent Zinkevich~\cite{zinkevich2003online} first showed the original regret bound which states that if the subgradient of $f_t$ is uniformly bounded by $G$ --- $\max_{y \in K} \max_{x \in \partial f(y)} \Vert x\Vert \leq G$ --- then we have
\[
\mathsf{Regret}(\mathcal{A}, T) \leq \left(\frac{\mathsf{diam}(K)^2 + G^2}{2}\right) \sqrt{T}
\]
where $\mathsf{diam}(K) = \max_{x, y \in K} d(x,y)$ is the diameter of a set.
These bounds were improved for $\mu$-strongly convex functions by Hazan~\cite{hazan2007logarithmic}, who showed that for $\mu$-strongly convex functions one has
\begin{equation}\label{eq:hazan-regret}
\mathsf{Regret}(\mathcal{A}, T) = O\left(\frac{G^2}{\mu} \log T\right)
\end{equation}
Finally, we note that for strongly convex and Lipschitz losses, the dynamic regret can be bounded as~\cite[Thm. 4]{besbes2015non}
\begin{equation}\label{eq:besbes-regret}
\mathsf{DRegret}(\mathcal{A}, T) = O\left(\frac{G^2}{\mu} \log T + \frac{G}{\mu} P_T\right)
\end{equation}
where $P_T$ is the so-called path length defined as
\[
P_T = \sum_{t=2}^{T} \Vert x^{\star}_t-x^{\star}_{t-1}\Vert_2
\]
where $x_t^{\star} = \argmin_{x\in K} f_t(x)$.
One proves this by noting that
\begin{align*}
\mathsf{DRegret}(\mathcal{A}, T) &= \sum_{t=1}^T \max_{x\in K} f_t(x) - \sum_{t=1}^T f_t(x_t) \\
&=  \sum_{t=1}^T \max_{x\in K} f_t(x) - \max_{x\in K} \sum_{t=1}^T f_t(x) + \max_{x\in K} \sum_{t=1}^T f_t(x) 
- \sum_{t=1}^T f_t(x) \\
& = \mathsf{Regret}(\mathcal{A}, T) + \left(\sum_{t=1}^T \max_{x\in K}  f_t(x) 
- \max_{x\in K}  \sum_{t=1}^T f_t(x)\right)
\end{align*}
Using the Lipschitz property and strong convexity, one can bound the paranthesized term by $O(P_t)$.

\section{Lending Protocols with a Single Borrowable Asset}\label{sec:single-dimensional-model}
DeFi lending protocols are overcollateralized protocols where borrowers have to deposit collateral worth more than what they are borrowing in another asset.
These systems have multiple agents and utilize different techniques to ensure solvency, which is the constraint that the assets lent by suppliers are always great than the liabilities held by the borrowers.
We describe a general model for online lending protocol revenue optimization.
While prior work (\eg~\cite{bertucci2024agents}) focuses on a stochastic control description of the process in continuous time, we will instead focus on modeling lending protocols as discrete-time processes.
Our goal is to incorporate the pooled and curation models in a single framework, and in many ways, the curation model is inherently discrete (as it involves curators bidding on interest rates).

\subsection{Agents}
In this paper, we will focus on three agents within lending protocols: suppliers, borrowers, and curators.
In practice, there are other agents, such as liquidators, who ensure that the protocol is solvent by buying collateral for defaulted loans from the protocol.
Since this paper focuses on revenue maximization, we are eliding the effect of liquidators and assuming that there is always a buyer of defaulted collateral.

\paragraph{Suppliers.}
Suppliers are agents who own crypto assets that they want to earn yield on.
They pool their assets within smart contracts that lend them out.
Assets that are utilized earn interest, while assets that are liquidated can have losses.
In the worst case, cascading liquidations can inure suppliers with large losses; however, in general, suppliers tend to be able to earn stable yields when loan quality is high.

\paragraph{Borrowers.}
Borrowers place collateral into a smart contract and borrow assets pooled together from suppliers.
They pay an interest rate to suppliers that depends on the size of their loan, the duration, and potentially the other loans that are demanded from the protocol after they open their loan.
If the interest payments grow larger than the collateral and/or if the value of the collateral falls below the value of the borrowed assets, the borrower is said to be in default.
Lending protocols have different mechanisms for liquidating defaulted loans, which can impact the pricing and/or revenue of a loan.
We assume a form of no-arbitrage (see~\S\ref{sec:lending-assumptions}) in that we assume that all liquidatable loans are immediately liquidated.

\paragraph{Curators.}
Curators are third-party actors who reallocate supplied assets to different markets.
Smart contracts are used to ensure that curators cannot manipulate or steal any of the supplied assets and can only migrate funds from one market to another.
If there are $M$ markets to which an asset $S$ can be supplied, a curator chooses a distribution $\pi \in \Delta^M$ and allocates $\pi_m S$ to market $m \in [M]$.
In practice, curators charge fees to suppliers as a function of the yield they earn, akin to a fund manager.
However, we ignore the utility of the curator in this document and assume that it is rational for them to maximize revenue for the suppliers\footnote{We note that there are numerous principal-agent incentive problems endemic to this setting, akin to those in liquid staking~\cite{tzinas2023principal}.}

\subsection{Single-Dimensional Model}
We start with a single-dimensional model, where there is one collateral asset and one borrowable asset, different from the collateral asset.
This models a single isolated market within Morpho~\cite{metamorpho}.
We assume that there is a time window $T \in \naturals$ that represents the maximum duration of a loan.

\paragraph{Supply Process.}
The supplied borrowable assets are grouped into $N \in \naturals$ vaults with capacity $S_n, n \in [N]$, which represent pools of the borrowable asset grouped together.
We also define the total capacity of the network as \[
S_{\text{total}} = \sum_{n \in [N]} S_n
\]
Generation one and two protocols corresponds to $N=1$ vaults, whereas Morpho-style protcools have $N > 1$ vaults.
Each vault has a strategy, defined as a value $\alpha_{n, t} \in [0, 1]$ that represents the amount of supply that is allocated to the market.
In the Aave market, we assume that $\alpha_{n, t} = 1$ for all $t$. 
At any given time, we define the active supply at time $t$, $S_t$ as
\[
S(\alpha, t) = \sum_{n \in [N]} \alpha_{n,t} S_n
\]
We can also define the revenue realized by supplier $n \in [N]$ as
\begin{equation}\label{eq:supplier-revenue}
R_n(t) = \left(\frac{\alpha_{n, t}S_n}{S(\alpha, t)}\right)R(t)
\end{equation}
where $R(t)$ is the total revenue earned by the system at time $t$.
We will define the revenue $R(t)$ on a per protocol basis later in the paper.

\paragraph{Demand Arrival Process.} 
The system evolves as a discrete-time stochastic process where at each time $t \in [T]$ the following actions are allowed:
\begin{itemize}
    \item Borrowers can open a loan $\ell_t$ of duration $\tau_t$
    \item Curators can reallocate supply 
\end{itemize}
The process $\ell_t \in \reals$ represents a single arriving or departing loan.
We say that a loan arrives if $\ell_t > 0$ and departs if $\ell_t < 0$.
$\tau_t \geq 0$ represents the duration of a loan.
That is, upon arrival at time $t$, the loan is active for $\tau_t \in \naturals \cup \{0\}$ steps.
In particular, this enforces the constraint $\ell_{t+\tau_t} = -\ell_t$.
We also enforce the constraint that if $\ell_t < 0$ then $\tau_t = 0$.
We note that the assumption that each $t \in [T]$ corresponds to a single action is not restrictive in a blockchain context, where loan operations are always executed sequentially.
Given a sequence of loans, we can define the active demand held by the protocol.
The active demand is simply defined as the sum of all loans that have arrived
\begin{equation}\label{eq:active-demand}
D(\{\ell_t\}, \{\tau_t\}, t) = \sum_{s=1}^t \ell_t = \sum_{s=1}^t \ones_{\ell_t \geq 0} \ones_{t < s + \tau(s)} \ell_t
\end{equation}
When the context is clear, we will abuse notation and write $D(t)$.

% \paragraph{Matching.}
% At each time $t \in [T]$, we match each arrival demand $q_t$ to a a supplier $i \in [n]$.
% We denote the matched supplier at time $t$ as $i_t$ and consider the sequence $I = (i_1, \ldots, i_T)$.
% For each $t \in [T]$, let $Q(i, t) = \sum_{s \leq t: i_s = i} q_t$, which represents the demand matched to the $i$th supply vault at time $t$.
% By construction, any valid matching $I$ satisfies $Q(i, t) \leq S_i$ for all $t \in [T]$.
% This states that the matched demand never exceeds the supply held at the $i$th vault.

\paragraph{Fixed Interest Pricing.}
We first consider a fixed price model, where upon each loan $\ell_t$ arriving, a price is quoted to them based on allocated supply.
We assume that each supplier first chooses $\alpha_{n, t}$ so that we can construct $S(\alpha, t)$.
Then a price $p_t = p_t(S(\alpha, t), D(t-1), \ell_t)$ is chosen given the new supply, the previous demand, and the new loan.
The price $p_t$ is the fixed price that the user pays over the duration $\tau(t)$.
This allows us to write the revenue $R^{\text{FI}}(\{p_t\}, t)$ as
\[
R^{\text{FI}}(\{p_t\}, t) = \sum_{s \leq t} p_t \tau(t) \ell_t \ones_{\ell_t > 0}
\]
This is the total income realized by suppliers given the prices $p_t$.
Note that the prices $p_t$ are chosen sequentially and do not know the future demand.
We will mainly consider rectilinear prices of the form
\[
p_t(S(\alpha, t), D(t-1), \ell_t) = \kappa \min\left(\frac{D(t-1) + \ell_t}{S(\alpha, t)}, 1\right) = \kappa \min\left(\frac{D(t)}{S(\alpha, t)}, 1\right)
\]
This price is linear in utilization $U(t) = \frac{D(t)}{S(\alpha, t)}$ until demand outweighs supply.
We enforce the constraint that no new loan can be added if $D(t) > S(\alpha, t)$.
This corresponds to saying that a lending protocol can only lend at most the supplied assets it has on hand.
We can represent this with a modified demand function:
\[
D(t) = \sum_{s=1}^t \ell_t \ones_{D(t-1) + \ell_t \leq S(\alpha, t)}
\]

Given that virtually all interest rate curves in DeFi are piecewise linear and convex, studying this function is sufficient to describe all live protocols.
For these prices, the fixed income revenue has the form
\[
R^{\text{FI}}(t) = \sum_{s \leq t} \kappa \frac{D(t)\ell_t \tau(t) \ones_{\ell_t \geq 0}}{S(\alpha, t)}
\]
Finally, we define the optimal fixed interest revenue in hindsight as
\[
R^{\text{FI}, \star}(\{\ell_t\},T) = \max_{p_1, \ldots, p_n} \sum_{s =1}^T p_i \tau(t)\ell_t \ones_{\ell_t \geq 0} 
\]
where the maximum is over all admissible price sequences.
Note that this revenue is optimal if we knew the demand $\ell_1, \ldots, \ell_T$ ahead of time, and hence $R^{\text{FI}}(T) \leq R^{\text{FI}, \star}(T)$.
We can consider the online pricing algorithm $\mathcal{A}^{\text{FI}}$ that offers the prices $p_t$.
For this algorithm, we have
\[
\mathsf{DRegret}(\mathcal{A}^{\text{FI}},T) = R^{\text{FI}, \star}(T) - R^{\text{FI}}(T)
\]
as we can think of each function $f_i(D) = p_i D$.

\paragraph{Variable Interest Pricing.}
Most lending protocols, including Aave, Compound, and Morpho, utilize variable interest rate pricing.
That is, the price a borrower pays can depend on the loans that arrive after opening a position.
One can view this as the protocol that quotes prices $p_1, \ldots, p_T$ at each time step, and the borrower pays the integrated cost over their duration.
Formally, we define the revenue at time $t$ to be
\[
R_t(p_1, \ldots, p_T) = \ones_{\ell_t > 0} \ell_t \left(\sum_{s=t}^{t+\tau(t)} p_s\right)
\]
where $\tau(t)$ is the duration of the arriving loan.
Note that the revenue process for variable interest rates is not adaptive and, in practice, stochastic control methods will fail since the revenue depend on future arrivals.
If we have linear pricing, the revenue becomes
\[
R_t = \ones_{\ell_t > 0} \kappa \ell_t \left(\sum_{s=t}^{t+\tau(t)} \frac{D(s)}{S(\alpha, s)}\right)
\]
We define the variable interest revenue as
\[
R^{VI}(t) = \sum_{s \leq t} R_s
\]
We can similarly define the optimal variable revenue as
\[
R^{\star}_t = \max_{p_1, \ldots, p_T} R_t(p_1, \ldots, p_T)
\]
and define the optimal variable revenue as $R^{VI, \star}(t) = \sum_{t \in [T]} R^{\star}_t$.
This allows us to compute the dynamic regret as
\[
\mathsf{DRegret}(\mathcal{A}^{VI}, T) = R^{VI, \star}(T) - R^{VI}(T)
\]

\paragraph{Pooled vs. Curated Models.}
In the above pricing descriptions, we assumed that there were $N$ curators each providing a strategy $\alpha_{n,t}$.
For the case of Generation 1 and 2 protocols, like Aave, where $N=1$, there is no strategy since there is a single pool.
We denote the variable interest rate algorithm by protocols as $\mathcal{A}^{\text{Pool}}$ and call them pooled strategies.

These protocols have infrequent updates of their allocations $\alpha_{n, t}$ through decentralized governance; when there is no change from governance, the supply can only change if users decide to withdraw assets.
Theoretically, if users who supply to pooled protocols were perfectly elastic to interest rate changes, they could replicate curated models.
Empirically, this does not occur, as most supply-side users tend to be passive liquidity providers~\cite{aave-ir-curve-changes} who update their holdings infrequently.
This means that one can think of pooled models as those where the supply is generally inelastic to rate changes.
As such, we describe the pooled model as a model in which the supply $S(\alpha, t)$ is fixed at all times $t \in [T]$.

On the other hand, curated models involve strategic, rational parties who earn performance fees based on how well they maximize revenue for suppliers.
This means that curated models can be thought of lending protocols with elastic supply that adapts to demand as it arrives.
Although suppliers can adapt their supply, they also have risks beyond those of standard lending protocols.
Curators have to manage inventory across pools and face opportunity and sourcing costs for misallocated inventory.
In some sense, curators have to solve a more complex version of the traditional newsvendor problem from inventory optimization~\cite{qin2011newsvendor}.

Our goal will be to compare the pooled model, with fixed supply, to the curator model that has adaptive supply modified by curators.
We will denote $\mathcal{A}^{\text{FI, Curator}}, \mathcal{A}^{\text{VI, Curator}}$ as the associated online algorithms that arise from curators competing to adjust the supply $S(\alpha, t)$.
We note that if all pooled depositors were strategic and not passive, one could replicate this strategy by having them all cooperate (but as mentioned, this does not occur in practice, necessitating curation).

\subsubsection{Examples}
We provide simple examples of how the Aave and Morpho models can differ in terms of their ability to maximize revenue.
Our examples quantify both the regret of the two models and the competitive ratio.
For simplicity, we consider the fixed interest rate model only but note that our examples can easily be extended to the variable case.

In this section, we have three main examples.
The first example demonstrates that pooled models can have $\Omega(T)$ regret (the worst possible regret for bounded functions), while on the same demand, the curation model has $O(1)$ regret.
This example, in particular, demonstrates that the results of~\S\ref{sec:results} that show that curation can cause $O(\log T)$ regret do not apply to the pooled model.
The second example demonstrates a similar effect, albeit with a competitive ratio.
The third example shows that the curation model degrades into the pooled model if there are only large-sized loans (relative to the amount of borrowable asset available) and the durations of the loans are long.

\paragraph{Example 1: Pooled models can have $\Omega(T)$ regret.}
We will describe examples that demonstrate that the Aave model can achieve $\Omega(T)$ regret and dynamic regret.
Consider both the pooled model with fixed supply $S = 1$ and the curated model with total supply $S_{\text{total}} = 1$.
At each time $t \in [T]$, a loan $\ell_t$ arrives and has fixed size $\frac{1}{T}$.
Each loan has a duration $T$, so we have demand $D(t) = \frac{t}{T}$.
For a rectilinear price, $p_t = \min\left(\frac{D(t)}{S(t)}, 1\right)$, the optimal supply is $S(t) = \frac{t}{T}$ so that $p_t = 1$ for all $t$.
This implies that we have
\[
R^{\text{FI},\star}(T) = \sum_{t\in [T]} p_t \tau(t) \ell_t  = T \sum_{t\in [T]} \frac{1}{T} = T
\]
On the other hand, for the Aave model we have
\[
p_t^{\text{Pool}} = \frac{D(t)}{S} = \frac{t}{T}
\]
This implies that the fixed interest revenue for the pooled model is
\[
R^{\text{FI}, \text{Pool}}(T) = \sum_{t\in[T]} p_t^{\text{Pool}} \tau(T) \ell_t = T \sum_{t\in [T]} \frac{t}{T^2} = T \cdot \frac{T(T+1)}{2T^2} = \frac{T}{2} + \frac{1}{2}
\]
since $\sum_{i=1}^n i = \frac{n(n+1)}{2} = \Theta(n^2)$.
This implies for $T$ sufficiently large
\[
R^{\text{FI}, \star}(T) - R^{\text{FI}, \text{Pool}}(T) = T - \frac{T}{2} - \frac{1}{2} = \frac{1}{2} (T-1) = \Omega(T) 
\]
which implies that the pooled model has $\Omega(T)$ (additive) regret.
However, in this example, notice that Aave receives a constant fraction of the optimal revenue since $\frac{R^{\text{FI},\star}(T)}{R^{\text{FI,Pool}}(T)} \approx 2$ as $T \rightarrow \infty$, which implies that the competitive ratio is constant.

\paragraph{Example 1: Curation achieves $O(1)$ regret.}
We assume the same setup of the last problem, except we replace the Aave interest model with the Morpho model.
We assume that each curator adjusts $\alpha_{n, t}$ using a local method (\eg~ordinary gradient descent) with step size $O\left(\frac{1}{t}\right)$~\cite{hazan2007logarithmic}.
This method converges to the optimal supply value $S^{\star}_t = D(t)$ with error $O\left(\frac{1}{t}\right)$.
This implies that $S(\alpha, t) = D(t) + O\left(\frac{1}{t}\right)$ via standard $O(\frac{1}{t}$) convergence results for gradient descent~\cite{hazan2007logarithmic}.
This implies that $p_t^{\text{Curator}}$ satisfies
\[
p_t^{\text{Curator}} = \frac{D(t)}{S(\alpha, t)} = \frac{D(t)}{D(t) + \frac{K}{t}} = \frac{1}{1+\frac{K}{D(t) t}} \geq 1 - \frac{K'}{D(t)t}
\]
for the appropriate choice of $K, K'$, where the last expression follows from the expansion of the geometric series.
This implies that we have
\[
R^{\text{FI, Curator}}(T) = \sum_{t\in[T]}p_t^{\text{Curator}} \tau(t) \ell_t \geq T \sum_{t\in [T]}\ell_t - \frac{K'\ell_t}{D(t) t} = R^{\text{FI}, \star}(T) - \sum_{t \in [T]}\frac{K}{t^2}
\]

Since $\sum_{t \in [T]} \frac{1}{t^2} = O(1)$, this implies that
\[
R^{\text{FI}, \star}(T) - R^{\text{FI, Curator}}(T) = O(1)
\]

\paragraph{Example 2: Pooled Model has a $O\left(\frac{1}{T}\right)$-competitive ratio.}
Now consider a sequence of loans $\ell_t$ that arrive and have duration $T-t$.
Each loan has size $\ell_t = \frac{1}{T^2}$ so that the demand satisfies $D(t) = \frac{t}{T^2}$.
As before, we have the optimal price $p^{\text{FI}, \star}_t = 1$ so that
\begin{equation}\label{eq:FI-star}
R^{\text{FI}, \star}(T) = \sum_{t \in [T]} p^{\text{FI}, \star}_t \tau(t) \ell_t = \sum_{t\in [T]} \frac{1}{T} = 1
\end{equation}
On the other hand, the Aave model has $p_t^{Pooled} = D(t)$ so we have
\[
R^{\text{FI, Pool}}(T) = \sum_{t\in [T]} D(t) \tau(t) \ell_t = \sum_{t\in [T]} \frac{t}{T^3} = \frac{T(T+1)}{2T^3} \leq \frac{1}{T}
\]
This implies that
\[
CR^{\text{Pool}}(T) = \frac{R^{\text{FI, Pool}}(T)}{R^{\text{FI}, \star}(T)} \leq \frac{1}{T}
\]
In particular, this means that the pooled model only realizes at most $\frac{1}{T}$ of the optimal revenue.

\paragraph{Example 2: Curation has $\Omega(1)$-competitive ratio.}
Note that the formula in~\eqref{eq:FI-star} also implies that we have $R^{\text{FI},\star}(T) \leq 1$.
Moreover we have $D(t) = \frac{t}{T^2} \leq \frac{1}{T}$.
This implies that we have
\[
p_t^{\text{Curator}} = \frac{D(t)}{D(t) + \frac{K}{t}} \geq \frac{D(t)}{\frac{1}{T} + \frac{K}{t}}
\]
This implies that we have a lower bound on Morpho revenue of
\begin{align*}
R^{\text{Curator}}(T) &= \sum_{t\in[T]} p_t^{\text{Curator}} \tau(t) \ell_t = \frac{1}{T} \sum_{t\in[T]}p_t^{\text{Curator}} \\
&> \frac{1}{T} \sum_{t\in [T]} 1 - \frac{K'}{D(t) t} = \frac{1}{T} \sum_{t\in [T]} 1 - \frac{K'T}{t^2}  
\end{align*}
proving that $R^{\text{Curator}}(T) = \Omega(1)$ and hence $CR^{\text{Curator}}(T) = \frac{R^{\text{FI, Curator}}(T)}{R^{\text{FI}, \star}(T)} = \Omega(1)$.

\paragraph{Example 3: Large loans cause both $\mathcal{A}^{Pool}$ and $\mathcal{A}^{Curator}$ to have a $\Omega(T)$ average regret.}

We now provide a simple example to show that if there are large loans (\ie~$\Omega(S))$ with long duration (\ie~$\Omega(T)$), then both the pooled and curation models have $\Omega(T)$ regret.
This example illustrates that our assumptions in the claims we will prove in~\S\ref{sec:results} are necessary to achieve low regret.
We consider a model where the pooled model has a supply $S = 1$ and the curation model has supply $S(\alpha, t) \leq S = 1, \forall t\in [T]$.
Our loan sequence is $\ell_1 = (1-\delta), \ell_t = \delta$ for all $t > 1$ and $\tau(t) = T$ for all $t \in [T]$.
This means that we have $D(t) = (1-\delta) + (t-1)\delta$.
This implies that the optimal revenue is
\[
R^{FI, \star} = \sum_{t\in [T]} p_t^{FI, \star} \tau(t) \ell_t = T\left((1-\delta) + (T-1) \delta\right) = T^2 \delta +  T(1-2\delta) = \Theta(T^2)
\]
On the other hand, both $R^{\text{Pool}}(T) = R^{\text{Curator}(T)} = T (1-\delta)^2 + T \delta = \Omega(T)$ as they can only satisfy the demand of $\ell_1, \ell_2$ before being at full utilization.
As such, both models receive $O(\frac{1}{T})$ of the optimal revenue in this case.

\section{Lending Protocols with Multiple Borrowable Assets}\label{sec:multidimensional-model}
Most lending markets provide a means to have multiple borrowing assets.
Users are allowed to deposit one of the $C \geq 1$ types of collateral and borrow $B \geq 1$ assets.
In Morpho, these markets are constructed by aggregating numerous isolated markets, whereas they are endemic to Euler.
One can view each curator $n \in [N]$ as having a supply of each borrowable asset, $b \in [B]$, $S^n_b$.
These assets are distributed in individual isolated markets where a user who offers collateral $c \in [C]$ can borrow the asset $b$.
For example, consider a borrowable asset that is USD.
A user who tenders Ethereum as collateral might be able to borrow less than a user who tenders Bitcoin, due to volatility and liquidity considerations.
The amount they can borrow for a given collateral type and the interest charged are determined by the amount of borrowable assets allocated to a particular collateral type.

\subsection{Background}
\paragraph{Allocation Matrix.}
The main difference between the single borrowable asset model is that now curators have to determine allocations to each collateral market.
We model this by having each curator $n$ have a matrix $A^{n}_{b,c}$ where $A^{n}_{b,c} \in [0, 1]$ is the percentage of their supply allocated to collateral type $c$.
For a curator with supply $S^n_b$ of the asset $b$, this means that they have allocated $S^n_b A^n_{b,c}$ units of asset $b$ to the market with collateral $c \in [C]$.
We naturally have the constraint $\sum_{c\in [C]} A^n_{b,c} = 1$, which implies that the matrices $A^n$ are column-stochastic.
Let $\mathcal{C}(B, C)$ be the set of column stochastic matrices of dimension $B \times C$.
As such, we call the matrix $A^n \in \mathcal{C}(B, C)$ the curator's allocation matrix $n$.
Finally, we assume that curators dynamically adjust their allocation matrix $A^n_{b,c}(t)$ in response to changes in demand and/or revenue.

\paragraph{Arrival Process.}
We assume that at each time $t \in [T]$, only a single action (open loan or close loan) occurs.
A multidimensional loan is defined as a tuple $(b_t, \ell_t, \tau(t)) \in [B] \times \reals^C \times \reals$.
We assume, as before, that at time $s = t + \tau(t)$, we have $\ell_s = -\ell_t$ (\ie~the loan is either repaid or liquidated in full after the duration).
The element $b_t$ is the type of the (single) borrowed asset that is lent, the vector $\ell_t \in \reals^C$ represents the amount borrowed from each collateral market, and $\tau(t)$ is the duration.

\paragraph{Supply, Demand, and Utilization.}
Consider a sequence of multidimensional loans, $\{(b_t, \ell_t, \tau(t))\}_{t\in [T]}$.
For each asset $b \in [B]$ we define the demand $D_b(t) \in \reals^C$ as
\[
D_b(t) = \sum_{s=1}^t \ell_s \ones_{b_s = b} \ones_{t < s+\tau(s)}
\]
This can be viewed as the set of unexpired loans who requested to borrow asset $b \in [B]$.
We denote the coordinate for the collateral $c \in[C]$ of this vector as $D_{b,c}(t)$.
For each borrowable asset $b\in [B]$, we define the supply of asset for collateral $c \in [C]$, $S_{b,c}(t)$, as
\[
S_{b, c}(A^1(t), \ldots,A^n(t)) = \sum_{n \in [N]} S_{b}^nA^n_{b,c}(t)
\]
Using these two definitions we can define the utilization for the pair $(b,c)$ as
\[
U_{b,c}(A^1(t), \ldots, A^n(t), t) = \min\left(\frac{D_{b,c}(t)}{S_{b,c}(A^1(t), \ldots, A^n(t), t)}, 1\right)
\]

\paragraph{Dynamics.}
Our model assumes that the dynamics of the system is the following:
\begin{enumerate}
    \item Curators adjust their allocations $A^n_{b,c}(t)$ from $A^{n}_{b,c}(t-1)$ via an online algorithm (\eg~Follow the Regularized Leader~\cite{hazan2016computational, hazan2022introduction})
    \item Loan $(b_t, \ell_t, \tau(t))$ arrives and supply and demand for each market are adjusted
\end{enumerate}

\subsection{Revenue and Regret}
In order to define the regret for $N$ curators competing in the multidimensional market, we will first need to consider the single curator market (\ie~a monopolist market).
For this market, we can more easily define revenue and regret.
After we define the monopolist's regret, we then define regret for the competitive curator model and related competitive to monopolist regret.

\paragraph{Monopolist Revenue.}
We will first define the revenue for $N=1$ (\eg~single curator).
This will allow us to define the optimal revenue in hindsight, which will be our benchmark for regret.
To do this, let $\kappa_{b,c} \in \reals_+$ be the elasticities (\eg~linear coefficients) for utilization.
Suppose first that we have a single static allocation matrix $A$ that does not change at each time $t \in [T]$. 
We can define the revenue for this function $R(A, T)$ as
\[
R(A, T) = \sum_{t \in [T]} \sum_{b \in [B]} \sum_{c \in [C]} \kappa_{b,c} U_{b,c}(A, t)
\]
where we assume that the supply used for the asset $b$ in utilization is equivalent to $\sum_{n \in [N]} S^n_{bc}$.
We define the optimal hindsight revenue as
\[
R^*(T) = \min_{A \in \mathcal{C}(B, C)} R(A, T)
\]

\paragraph{Monopolist Regret.}
Next, we assume that there is a single curator $(n=1)$ but that they update their allocations $A_1, \ldots, A_T$ sequentially.
In this case, we can define the revenue
\[
R(A_1, \ldots, A_T) = \sum_{t \in [T]} \sum_{b \in [B]} \sum_{c \in [C]} \kappa_{b,c} U_{b,c}(A_t, t)
\]
We can now define the regret for the monopolist as
\[
\mathsf{Regret}^{Monopolist}(T) = R(A_1, \ldots, A_T) - R^*(T)
\]
We will show in~\S\ref{subsec:multidimensional-results} that $\mathsf{Regret}^{Monopolist}(T) = O(B \log C \log T)$ when the utilization stays in the strongly convex region bounded away from 1.

\paragraph{Curatorial Regret.}
We define the revenue with $n$ curators as
\[
R(\{A^n_t\}_{n \in [N], t\in[T]}) = \sum_{t \in [T]} \sum_{b \in [B]} \sum_{c \in [C]} \kappa_{b,c} U_{b,c}(A^1(t), \ldots, A^n(t), t)
\]
and the regret
\[
\mathsf{Regret}^{Curator}(\{A^n_t\}_{n \in [N], t\in[T]}) = R(\{A^n_t\}_{n \in [N], t\in[T]})  - R^{\star}(T)
\]
We first note that if $\hat{A}(t) = \sum_{n \in [N]} A^n(t) \in \mathcal{C}(B, C)$ is the aggregate allocation matrix and if $\Vert \hat{A}(t) - A^{\star}\Vert_2 = O\left(\frac{\log t}{t}\right)$, then $\mathsf{Regret}^{Curator}(T) = O(B \log C (\log T)^2)$.
This implies that as long as the agents are able to learn an approximate allocation matrix that approaches the optimal (in a no-regret sense), then one only pays an excess cost of $O(\log T)$ over the monopolist.

% We provide an example here that shows that the regret of the pooled model and the RFQ model can differ dramatically.
% In this model, the supply provided by the RFQ model can adjust to demand and duration, allowing revenue to be approximately optimized.
% On the other hand, the pooled mode is stuck with a fixed supply and cannot adjust its pricing fast enough to deal with duration.

% Suppose $T = k^2$ for some $k \in \naturals$.
% Let the demand sequence be 
% \[
% q_s = q (\ones_{s \equiv 0 \text{ mod } 2k} - \ones_{s \equiv k \text{mod 2k}})
% \]
% This means that a loan is active for $k = \sqrt{T}$ timesteps and that there are no overlapping loans.
% There is also exactly $q$ units of borrow at each timestep.
% Imagine that we have a pooled model with $S = \Omega(T)$ units and an RFQ model that satisfies $S_t(\alpha) = \sqrt{T}$.
% For this demand, the price quoted by the pooled model, at all times is $\frac{kq}{S} \leq \frac{kq}{T}$ whereas the RFQ model quotes $\frac{kq}{\sqrt{T}}$. 
% The revenue earned by the pooled model is
% \[
% R^{\text{Pool}}(T) = \frac{\kappa q}{S} T = O(1)
% \]
% and the revenue earned by the RFQ model is
% \[
% R^{RFQ}(T) = \frac{\kappa q}{\sqrt{T}} T = \kappa q \sqrt{T} = \Omega(\sqrt{T})  
% \]
% This shows that the gap between the regret for the RFQ system and the pooled system is at least $\sqrt{T}$ for this scenario.
% Moreover, this suggests that the gap between the pooled and RFQ model depends on the expected duration of a loan (\ie~the time that the loan is active).
\section{Main Results}\label{sec:results}
In this section, we provide our main results, which informally say the following:
\begin{itemize}
    \item For a single borrowable asset, if the loan sizes and durations are sufficiently small, then curation achieves $O(\log T)$ regret
    \item For $B$ borrowable assets and $C$ collateral assets, if the loan sizes and durations are sufficiently small, then curation achieves $O(B \log C \log T)$ regret
    \item For sufficiently large loans and durations, neither the pooled nor curation models can achieve better than $\Omega(T)$ regret 
\end{itemize}

\subsection{Assumptions}\label{sec:lending-assumptions}
We first lay out the assumptions that we use in our results in the following sections.
We will also prove some basic properties about these assumptions which will simplify the proofs of our main results.

\paragraph{Bounded (Compact) Supply.}
Given that the utilization of a lending protocol is inversely proportional to the assets supplied, our first assumption is that all markets have a bounded amount of supply.
In practice, virtually all lending protocols enforce this at the protocol level by adding minimum and maximum supply caps.
\begin{assumption}\label{assm:min-supply}
    All markets have $S(\alpha, t) \in [S_{\min}, S_{\max}]$ with $S_{\min} > 0$.
\end{assumption}

\paragraph{Minimum Demand.}
We also assume that there is some minimum amount of demand.
In practice, this also ends up being true as, in practice, there is always demand for looping strategies (cf.~\cite{what-is-looping}).
\begin{assumption}\label{assm:min-demand}
All lending markets have a minimum demand $D_{\min} > 0$ with $D(t) \geq D_{\min}$ for all $t \in [T]$
\end{assumption}

\paragraph{Bounded Loan Increments.}
A natural constraint on the loan process is to ensure that it is `smooth in a strong enough sense to bound changes in revenue.
At the same time, due to the discrete nature of our system, we also want to allow for infrequent but large jumps.
We do this with a subexponential bounded increment assumption:
\begin{assumption}\label{assm:bounded-increment}
Given a loan process $(\ell_s, \tau(s))_{s\in[T]}$, we say that it satisfies the subexponential bounded increment assumption with parameter $\Delta > 0$ if there exists $K > 0$ such that
\[
\Prob[|\ell_s - \ell_{s-1}| > \Delta] \leq e^{-K\Delta}
\]
holds for all $s \in [T]$
\end{assumption}
We note that the subexponential tail of this bound could be relaxed to a fatter tailed distribution, but we elide this to simplify our results.

\paragraph{Reset Condition.}
Our second assumption is a reset condition.
This can be thought of as a formal way of saying that the integrated demand (\eg~sum of demand over time) is not too large relative to the total of assets supplied that can be borrowed.
Another way of looking at this is saying that (relatively) large loans are paid off quickly.

\begin{assumption}\label{assm:reset-condition}
Given a loan process $(l_s, \tau(s))_{s\in[T]}$, we say that it satisfies the $\epsilon$-reset condition if for all $\epsilon' \geq \epsilon$ we have
\[
\Prob\left[\ell_s \tau(s) > \epsilon' \sum_{n\in[N]} S_n\right] \leq e^{-\Omega(\epsilon')}
\]
\end{assumption}
One can view $\ell_s \tau(s)$ as the total borrowed asset over time, and this condition says that the probability that it is a constant fraction of the total supplied decays exponentially.
We note that this can be relaxed to fatter-tailed distributions, but elide this complexity to simplify the proof.

\paragraph{Strategic Curator Competition implies High Utilization.}
We model the $N$ curators as strategic, rational competitors who earn pro-rata rewards for assets utilized and face costs for unutilized assets.
In particular, we think curators as inventory managers, much like in the classical newsvendor problem~\cite{qin2011newsvendor, petruzzi1999pricing}.
More formally, we assume that each strategic curator receives a revenue proportional to their liquidity, $R_t \propto \frac{\alpha_{n, t} S_n}{S(\alpha_t, t)}$.
For the unused liquidity --- $(1-\alpha_{n,t})S_n$ --- we say that each curator faces costs $C_n(1-\alpha_{n, t})$.
Our main assumption provides conditions on costs $C_n$ to ensure that a constant fraction of the total supply is always available.

\begin{assumption}\label{assm:curator-costs}
For each curator $n \in [N]$, the following conditions hold:
\begin{enumerate}
\item \textbf{Net Profit.} Curator has a net profit $\pi_t(\alpha) = R_n(\alpha) - C_n(1-\alpha_{n,t})$ where $R_n(\alpha)$ is defined as
\[
R_n(\alpha) = \frac{\alpha_{n,t}S_n}{\sum_{n\in[N]} \alpha_{n,t} S_n} R(t)
\]
where $R(t)$ is the total protocol revenue at time $t$ and $C_n : [0,1] \rightarrow \reals_+$ is a convex, differentiable, and increasing cost function with $C_n(0) = 0$.
\item \textbf{Weak Rationality.} Curator are weakly rational in that they try to (locally) maximize their profit
\item \textbf{Existence of Low Cost Curation.} There exists $c^{\star} \in (0,1]$ and $A \subset [N]$ with $|A| \geq p N$ for some $p > 0$ such that for all $n \in A$ we have $C'_n(0) \leq c^{\star} S_n$
\item \textbf{Minimum Protocol Revenue.} For all rounds $t \in [T]$, we have $R(t) \geq c^{\star} S(\alpha, t)$
\end{enumerate}
\end{assumption}

In Appendix~\ref{app:supply-claim}, we show the following claim:
\begin{claim}\label{claim:supply-claim}
    Suppose that Assumption~\ref{assm:curator-costs} holds and that each supplier plays no-regret dynamics (\eg~ordinary gradient descent) to choose $\alpha_{n, t}$.
    Then there exists $c > 0$ such that $S(\alpha, t) \geq c \sum_{n\in[N]} S_n$ for all $t \in [T]$
\end{claim}

\paragraph{Variable Interest Rate Concentration.}
In order to analyze the variable rate model, we need to constrain the deviation of the interest rate over the duration of the typical loan.
We do this by restricting how much the variable rate loan deviates from the fixed rate price.
Let $\sigma_p > 0$ be the variance of the price process in the variable rate model.
Our assumption is that the variance and duration of the loan control the overall revenue.
\begin{assumption}\label{assm:variable-rate}
Let $\sigma_p$ be the variance of the price process. We say that the variance bounded rate assumption holds if for all $t\in [T]$ the follding tail bound holds
\[
\Prob\left[\left|\sum_{s=t}^{t+\tau(t)} p_t - \Expect[\tau(t)] p_t \right| > (1+\epsilon) \sigma_p \sqrt{\tau(t)} \right] \leq e^{-\Omega(\epsilon)}
\]
\end{assumption}

\paragraph{Minimum Allocation Amount.}
For the multiple borrowable asset problem, for the problem to stay strongly convex, we need to avoid the boundary of the $C$-dimensional probability simplex.
A simple constraint to enforce that is to assume that there exists a constant $a > 0$ such that every allocation $A^n_{bc}(t) > a$ for all $t \in [T], b \in [B], c \in [C]$:
\begin{assumption}\label{assm:min-alloc}
We say that the minimum allocation assumption is valid if there exists $a > 0$ such that for all $t \in [T], b\in [B], c\in[C]$, we have $A^n_{bc}(t) > a$.
\end{assumption}
Note that this implies that $S_b(t) = \sum_{c\in[C]} S_{b,c}(t) > a C \sum_{n \in [N]} S^n_b > 0$, \ie~that there is a minimum amount of supply for every borrowable asset.
We utilize this within the proof of Claim~\ref{claim:monopolist}.

\paragraph{Maximum Elasticity}
Our final assumption is that in the multidimensional model we need to ensure that there is a uniform bound on the elasticities $\kappa_{b,c}$.
\begin{assumption}\label{assm:max-elasticity}
We assume that there is $K > 0$ such that $\max_{b \in [B], c\in [C]} \kappa_{b,c} \leq K$.
\end{assumption}

\subsection{Single Borrowable Asset}
We first consider the fixed interest rate model, as we use these results to bound the variable interest rate (multiple rounds) model.
We define the environment $\mathcal{E}(n, T)$ as a setting with observation time $T \in \naturals$, $N$ curators with maximum supply $S_n, n \in [N]$ and a loan process $(\ell_s,\tau(s)), s\in [T]$.

\paragraph{Fixed Interest Rate.}
We make the following claim about the regret for $\mathcal{A}^{Curation}$ that we prove in Appendix~\ref{app:single-borrow-regret}
\begin{claim}\label{claim:single-borrow-regret}
Suppose that we have an instance of $\mathcal{E}(n, T)$ and that assumptions~\ref{assm:min-supply},~\ref{assm:min-demand},~\ref{assm:bounded-increment},~\ref{assm:reset-condition}, and~\ref{assm:curator-costs} hold.
Then, with probability greater than $1-\Omega\left(\frac{1}{T}\right)$, we have
\[
\mathsf{Regret}(\mathcal{A}^{Curation}, T) \leq O\left((\log T)^2\right)
\]
\end{claim}
\noindent 
We note that if we make a strong assumption on the distributions in our assumptions, \eg~make them Subgaussian instead of Subexponential, then we get $O(\log T)$.
However, since the empirical distributions do not appear to be sub-Gaussian, we used weaker assumptions~\cite{aave-ir-curve-changes, aave-ir-curve-changes-2}.
If we add some further assumptions on the tail of the distribution of loan size and duration, we can get sublinear dynamic regret. 

\begin{claim}\label{claim:single-borrow-dregret}
If in addition to the hypotheses of Claim~\ref{claim:single-borrow-regret}, we also have Assumption~\ref{assm:bounded-increment} hold with $\Delta = o_T(1)$, then we have
\[
\mathsf{DRegret}(\mathcal{A}^{Curation}, T) \leq O\left( (\log T)^2 + o(T) \right)
\]
\end{claim}
\noindent This follows from the proof of Claim~\ref{claim:single-borrow-regret} and noting that that $\Delta = o_T(1)$ implies $P_t \leq T \Delta = o(T)$.

\paragraph{Variable Interest Rate Model.}
For the variable interest rate model, we have a similar result to Claim~\ref{claim:single-borrow-regret}:
\begin{claim}\label{claim:single-variable-regret}
Suppose that we have an instance of $\mathcal{E}(n, T)$ and that the assumptions~\ref{assm:min-supply},~\ref{assm:min-demand},~\ref{assm:bounded-increment},~\ref{assm:reset-condition},~\ref{assm:curator-costs}, and~\ref{assm:variable-rate} hold.
Then with a probability greater than $1-\Omega(\frac{1}{T})$, we have
\[
\mathsf{Regret}(\mathcal{A}^{Curation}, T)\leq O(\Expect[\tau(t)] \sigma_p^2 (\log T)^4)
\]
\end{claim}
In particular, this implies that if the average duration of the loans satisfies $\Expect[\tau(t)] = O(T^c)$ for $c < 1$, then we have sublinear regret.
We note that in practice, curators influence the volatility price process $\sigma_p$, which is an effect we ignore here.

\subsection{Multiple Borrowable Assets}\label{subsec:multidimensional-results}
Let $\mathcal{E}(N, B,C,T)$ be an environment with observation time $T$, $B$ borrowable assets, $C$ collateral assets, and $S^n_b \in \reals_+$ representing the supply held by the curators.
For simplicity, we only consider fixed-rate loans, but note that our proofs can easily generalize.

\begin{claim}\label{claim:monopolist}
    Suppose that we are in $\mathcal{E}(N, B, C, T)$ and Assumptions~\ref{assm:min-supply},~\ref{assm:min-demand},~\ref{assm:bounded-increment},~\ref{assm:reset-condition},~\ref{assm:curator-costs},~\ref{assm:min-alloc},~and~\ref{assm:max-elasticity} hold.
    Then with probability $1-\Omega(\frac{1}{T})$:
    \[
    \mathsf{Regret}(\mathcal{A}^{Monopolist}, T) = O(B \log C (\log T)^2)
    \]
\end{claim}
This is proved in Appendix~\ref{app:monopolist}.
We note that the $\log C$ factor comes from the fact that this can be viewed as $B$ separate learning problems on the $C$-dimensional simplex. 
Online learning on the simplex via mirror descent with a logarithmic barrier is known to achieve $O(\log C \log T)$ regret~\cite{hazan2016online}.

\begin{claim}\label{claim:curatorial}
    Suppose that we are in $\mathcal{E}(N, B, C, T)$ and the assumption of the previous claim holds.
    If, in addition, $\Vert \hat{A}(t) - A^{\star}\Vert_2 = O\left(\frac{\log t}{t}\right)$, then \[
    \mathsf{Regret}(\mathcal{A}^{Curatorial}, T) = O(B \log C (\log T)^3)
    \]
\end{claim}

\section{Conclusions and Future Directions}
In this paper, we formulate a model for lending protocols as online algorithms, generalizing prior models that view lending protocols via stochastic control.
We demonstrated that under mild assumptions about the maximum loan size and duration of a loan, adaptive supply mechanisms perform exponentially better in terms of regret than fixed supply mechanisms.
This large difference in regret directly follows from rational curators being incentivized to keep the revenue curve in the region where it is strongly convex.
Our results demonstrate that dynamic pricing (through adapting supply) can dramatically improve protocol revenue.

Our model can be extended in a number of ways.
First, we note that we assume that there is no feedback between interest rates and lending demand.
However, looping strategies, which are common in both Aave and Morpho, represent demand that adjusts in response to posted prices.
There are a number of ways to add feedback to our model and achieve similar regret and dynamic regret bounds (\cf~\cite{baby2022optimal}).
Moreover, our model does not include any notion of congestion costs amongst curators.
Given the known results of congestion costs that impact welfare in decentralized trading~\cite{chitra2024analysis, yuki-paper}, one would need to model this to extend our revenue results to welfare.

\bibliographystyle{ACM-Reference-Format}
\bibliography{bib.bib} 

\appendix

\section{Proof of Claim~\ref{claim:supply-claim}}\label{app:supply-claim}
From the definition of revenue (equation~\eqref{eq:supplier-revenue}), the supplier solves the optimization problem
\[
\alpha_{n,t} = \argmax_{\alpha \in (0, 1]} \frac{\alpha S_n}{\alpha S_n + \sum_{n'\neq n} \alpha_{n',t}S_{n'}}
\]
This function is strictly increasing in $\alpha$, which implies that the supplier's best response is $\alpha = 1$ and that $\alpha_{n, t} = 1$ for all $n\in[N]$ is a Nash equilibrium.
This implies that without costs, no-regret dynamics for the players would converge to $\alpha_{n,t} \rightarrow 1$ as $t \rightarrow \infty$, which implies $S(t) \rightarrow \sum_{n \in [N]} S_n$.

Now suppose that there exist positive costs $C_n$ for each curator that satisfy Assumption~\ref{assm:curator-costs}.
Let $A \subset [N]$, $p$ be the low cost set and density of this set, respectively.
For any $n \in A$, we first claim that $C'_n(0)$ is smaller than the marginal revenue $R'_n(0)$.
Let $S_{-n}(\alpha_t) = \sum_{m \neq n} \alpha_{m, t} S_m$.
Then we can write
\[
R_n(\alpha) = \frac{\alpha_{n,t} S_n}{\alpha_{n,t} S_n + S_{-m}(\alpha)}
\]
and $R'_n(0) = \frac{S_n}{S_{-m}(\alpha)} R(t) \geq c^{\star} S_n \geq C'_n(0)$.
This implies that $\alpha_{n,t} = 1$ is the best response for any member of $A$.
Therefore, at least $pN$ players are playing $\alpha_{n, t} = 1$ in equilibrium.
Since this is a concave pro-rata game, no-regret dynamics (\eg~OGD)converges at a rate $O\left(\frac{\log t}{t}\right)$, so that we will have $S(\alpha, t) \geq c\sum_{n \in [N]}S_n - \Theta\left(\frac{\log t}{t}\right)$
\section{Proof of Claim~\ref{claim:single-borrow-regret}}\label{app:single-borrow-regret}
Since Assumption~\ref{assm:curator-costs} holds, Claim~\ref{claim:supply-claim} implies that $S(\alpha, t) \geq c S_{\text{total}}$ for all $t \in [T]$ for $c > 0$.
We next show that with probability greater than $1-\Omega(\frac{1}{T})$, there exists $\epsilon > 0$ such that $D(t) \leq S(\alpha, T) - \epsilon$.
Assumption~\ref{assm:reset-condition} implies that with probability at least $1-e^{-c}$ we have $\ell_t \tau(t) \leq \epsilon S_{\text{total}} \leq \frac{\epsilon}{c} S(\alpha, t)$.
Let $\ell_{\max} = \max_{t\in[T]} \ell_t, \tau_{\max} = \max_{t \in [T]} \tau(t)$.
Using the tail bound in Assumption~\ref{assm:reset-condition} and the union bound, we have
\[
\Prob[\ell_{\max} \tau_{\max} > \frac{\epsilon}{c} S(\alpha, t)] \leq \Prob[\ell_{\max} \tau_{\max} > \epsilon S_{\text{total}}] \leq T \Prob[\ell_t \tau(t) > \epsilon S_{\text{total}}] \leq Te^{-\Omega(\epsilon)}
\]
Note that the usage of the union bound here implicitly utilizes Assumption~\ref{assm:bounded-increment}.
By letting $\epsilon = O(\log T)$, we get that this probability is $O\left(\frac{1}{T}\right)$.
Note that by construction $D(t) \leq \ell_{\max}\tau_{\max}$, since only $\tau_{\max}$ loans can be simultaneously active at any time.
Hence, we have $D(t) \leq \frac{\epsilon}{c} S(\alpha, t)$ with probability $1-\Omega(\frac{1}{T})$ as claimed.

Now we claim that the revenue $R^{Curation}(T)$ is strongly convex when $D(t) \leq \frac{\epsilon}{c} S(\alpha, t)$.
Define the loss function $L_t(S(\alpha, t)) = -R^{Curation}(t) = - \kappa \frac{D(t)\tau(t) \ell_t}{S(\alpha, t)}$.
This function has $L''_t(S(\alpha, t)) = -\kappa \frac{2D(t)\tau(t)\ell_t}{S(\alpha, t)^3} \leq -\frac{2\kappa D_{\max}^2 \tau(t)}{S_{\max}^3}$.
Using our tail bound, we have for $c < 1$
\begin{align*}
\Prob\left[\tau_{\max} > \frac{2}{K}\frac{S_{\text{total}}}{D_{\min}} \log T\right] &= T\Prob\left[\tau(t) > \frac{\epsilon S_{\text{total}}}{\ell_{t}}\right] \\
&\leq T\Prob\left[\tau(t) > \frac{\epsilon S_{\text{total}}}{D_{\min}}\right] \\
&\leq Te^{-2 \log T} = O\left(\frac{1}{T}\right)
\end{align*}
Tis implies that as $T \rightarrow \infty$, with high probability greater than $1-\frac{1}{T}$, we have $L''(S(\alpha, t) \leq - \frac{4\kappa D^2_{\max} \log T}{K D_{\min} S^2_{\max}} \leq \frac{4\kappa}{K D_{\min}} \log T$.
This implies that $\alpha \geq \frac{4\kappa \log T}{D_{\min}}$.
Similarly, note that with high probability, we have $G \leq \max_{S \in [S_{\min}, S_{\max}]} |L'_t(S)| \leq \kappa \frac{D_{\max}^2 \tau(t)}{S_{\min}^2} \leq  \kappa \frac{2 D_{\max}^2 S_{\text{total}}}{K D_{\min} S_{\min}^2} \log T$.
Therefore choosing constants appropriately and union bounding over probabilities, we have with probability at least $1-\Omega(\frac{1}{T})$,
\[
\frac{G^2}{\alpha} \leq \frac{4\kappa D^2_{\max} S^3_{\max}}{D_{\min} S^4_{\min}} \log T = O(\log T)
\]
Using eq.~\eqref{eq:hazan-regret}, this gives the answer.
\section{Proof of Claim~\ref{claim:single-variable-regret}}
Under Assumption~\ref{assm:variable-rate}, we have
$R_t = \ell_t \left(\sum_{s=t}^{t+\tau(t)}p_t\right) \leq \Expect(\tau(t))p_t + (1+\epsilon)\sigma_p\sqrt{\tau(t)}$ with high probability.
This implies that $G = O(\sigma_p\sqrt{\tau(t)} \log T)$ with probability $1-\Omega(\frac{1}{T})$.
Following the proof of the previous claim, this implies that the variable regret is bounded by $\Expect[\tau(t)]\sigma_p^2 (\log T)^2$ times the fixed regret.
\section{Proof of Claim~\ref{claim:monopolist}}\label{app:monopolist}

We first define a per-asset, per-collateral loss.
For a borrowable asset $b$ and collateral $c$, when the system is in the unsaturated region (i.e. $D_{b,c}(t) < S_{b,c}(A, t)$), we have the revenue
    \[
    R_{b,c}(A_{b,c}(t), t) = k_{b,c}\, \frac{D_{b,c}(t)}{S_{b,c}(A, t)},
    \]
Define $L_{bc}(x, t) = -R_{b,c}(x, T)$.
We show that the loss is strongly concave in this region.
Note that since $A_{b,c}(t) > a$ and there is a minimum supply, the denominator is in the strongly convex region of the function $\frac{1}{x}$.
As before, we now need to bound the gradient.
The gradient satisfies 
    \[
    L_{b,c,t}'(x) = \frac{k_{b,c}\,D_{b,c}(t)}{S_{b,c}\,x^2}.
    \]
    On $x\in [a,1]$, this is at most
    \[
    G_{b,c} = \frac{k_{b,c}\,d_{\max}}{S_{b,c}\,a^2},
    \]
For each $(b,c)$, using standard mirror descent on the $C$--dimensional simplex~\cite{hazan2007logarithmic}, we have a regret bound:
    \[
    \mathsf{Regret}_{b,c}(T) \le \frac{G_{b,c}^2}{\alpha_{b,c}} \log T.
    \]
Plugging in the values, we obtain
    \[
    \mathsf{Regret}_{b,c}(T) \le \frac{\left(\frac{k_{b,c}\,d_{\max}}{S_{b,c}\,a^2}\right)^2}{\frac{2k_{b,c}\,d_{\min}}{a^3\,S_{b,c}}} \ln T 
    = \frac{k_{b,c}\,d_{\max}^2}{2a\,d_{\min}\,S_{b,c}} \log T.
    \]
Naively, this would imply a $O(C \log T)$ bound, but we note that expert models (\eg~mirror descent plus a logarithmic barrier) can move this to $\log C$~\cite{hazan2016online}.
This implies a per borrow asset regret, $\mathsf{Reg}_b(T)$ of
    \[
    \mathsf{Reg}_b(T) = O\Bigl(\frac{K_b\,d_{\max}^2}{2a\,d_{\min}\,S_{b}} \ln C\,\log T\Bigr).
    \]
where $K_b = \max_{c \in [C]} k_{b,c}$, $S_b = \min_{c\in [C]} S_{b,c}$.
If we sum over $b \in [B]$, this gives the final bound
\[
\mathsf{Regret}(\mathcal{A}^{Monopolist}, T) = O\Bigl(\sum_{b=1}^B \frac{K_b\,d_{\max}^2}{2a\,d_{\min}\,S_{\text{total}}[b]} \ln C\,\ln T\Bigr) 
\]
Finally we assume that for all $b$, $K_b\le K$ and $S_{b}\ge S_0$, then
 \[
    \mathsf{Regret}(\mathcal{A}^{Monopolist}, T) \le O\!\Bigl(\frac{B\,K\,d_{\max}^2}{2a\,d_{\min}\,S_0} \log C\,\log T\Bigr) = O(B \log C \log T)
\]
as claimed.
The extra logarithmic factor in the claim comes from the probabilistic assumptions made (\eg~subexponential tails)

\section{Proof of Claim~\ref{claim:curatorial}}

Assume that the aggregate allocation at time $t$ is 
\[
\hat{A}(t) = \sum_{k=1}^K A^k_{bc}(t)
\]
and that the convergence error of the aggregate allocation to the optimal static allocation $A^*(t)$ satisfies
\[
\|\hat{A}(t)-A^*(t)\| \le \frac{C_1 \ln t}{t},
\]
for some constant $C_1 > 0$. Then the additional per-agent revenue loss is proportional to the allocation error, say, with Lipschitz constant $L$.
Note that $L < \infty$ since we assume the supply is bounded.
Therefore, the extra loss per round is at most
\[
L \cdot \frac{C_1 \ln t}{t}.
\]
Summing over $t=1$ to $T$, we obtain
\[
\sum_{t=1}^T \frac{C_1 L \ln t}{t} \le C_1 L \cdot O\bigl((\ln T)^2\bigr).
\]
Thus, the per-agent (or $K$-agent) regret is bounded by
\[
\mathsf{Regret}(\mathcal{A}^{Curator}, T) = O\Bigl(B\,\ln C\, (\ln T)^2\Bigr).
\]
The extra $O(\log T)$ term comes from the subexponential bounds.

\end{document}